\let\useblackboard=\iftrue
%
%
\newfam\black
\input harvmac.tex
\noblackbox
\def\Title#1#2{\rightline{#1}
\ifx\answ\bigans\nopagenumbers\pageno0\vskip1in%
\baselineskip 15pt plus 1pt minus 1pt
\else
\def\listrefs{\footatend\vskip 1in\immediate\closeout\rfile\writestoppt
\baselineskip=14pt\centerline{{\bf References}}\bigskip{\frenchspacing%
\parindent=20pt\escapechar=` \input
refs.tmp\vfill\eject}\nonfrenchspacing}
\pageno1\vskip.8in\fi \centerline{\titlefont #2}\vskip .5in}

\ifx\answ\bigans\def\tcbreak#1{}\else\def\tcbreak#1{\cr&{#1}}\fi
\useblackboard
\message{If you do not have msbm (blackboard bold) fonts,}
\message{change the option at the top of the tex file.}

\font\blackboard=msbm10 scaled \magstep1
\font\blackboards=msbm7
\font\blackboardss=msbm5
\textfont\black=\blackboard
\scriptfont\black=\blackboards
\scriptscriptfont\black=\blackboardss

\else

\fi
%
\def\yboxit#1#2{\vbox{\hrule height #1 \hbox{\vrule width #1
\vbox{#2}\vrule width #1 }\hrule height #1 }}
\def\fillbox#1{\hbox to #1{\vbox to #1{\vfil}\hfil}}
\def\ybox{{\lower 1.3pt \yboxit{0.4pt}{\fillbox{8pt}}\hskip-0.2pt}}

\def\comments#1{}

\def\half{{1\over 2}}
\def\Tr{{{\rm Tr\  }}}

\def\CN{{\cal N}}

\def\II{\relax{I\kern-.07em I}}
\def\IIA{{\II}A}

\def\inbar{\,\vrule height1.5ex width.4pt depth0pt}
\def\IZ{\relax\ifmmode\mathchoice
{\hbox{\cmss Z\kern-.4em Z}}{\hbox{\cmss Z\kern-.4em Z}}
{\lower.9pt\hbox{\cmsss Z\kern-.4em Z}}
{\lower1.2pt\hbox{\cmsss Z\kern-.4em Z}}\else{\cmss Z\kern-.4em
Z}\fi}
\def\IB{\relax{\rm I\kern-.18em B}}
\def\IC{{\relax\hbox{$\inbar\kern-.3em{\rm C}$}}}
\def\ID{\relax{\rm I\kern-.18em D}}
\def\IE{\relax{\rm I\kern-.18em E}}
\def\IF{\relax{\rm I\kern-.18em F}}
\def\IG{\relax\hbox{$\inbar\kern-.3em{\rm G}$}}
\def\IGa{\relax\hbox{${\rm I}\kern-.18em\Gamma$}}
\def\IH{\relax{\rm I\kern-.18em H}}
\def\IK{\relax{\rm I\kern-.18em K}}
\def\IP{\relax{\rm I\kern-.18em P}}

\font\cmss=cmss10 \font\cmsss=cmss10 at 7pt
\def\IR{\relax{\rm I\kern-.18em R}}

\def\sdtimes{\mathbin{\hbox{\hskip2pt\vrule
height 4.1pt depth -.3pt width .25pt\hskip-2pt$\times$}}}
\def\Tr{\rm Tr}

\def\BR{\IR}
\def\BZ{\IZ}
\def\BR{\IR}

\def\maphor#1{\smash{ \mathop{\longleftrightarrow}\limits^{#1}}}
\def\mapver#1{\Big\updownarrow
\rlap{$\vcenter{\hbox{$\scriptstyle#1$}}$}}

%
%

\def\NP{{\it Nucl. Phys.\ }}
\def\AP{{\it Ann. Phys.\ }}
\def\PL{{\it Phys. Lett.\ }}
\def\PR{{\it Phys. Rev.\ }}
\def\PRL{{\it Phys. Rev. Lett.\ }}

\def\JMP{{\it J. Math. Phys.\ }}

\def\IJMP{{\it Int. Jour. Mod. Phys.\ }}
\def\Mod{{\it Mod. Phys. Lett.\ }}

\Title{ \vbox{\baselineskip12pt\hbox{hep-th/9704041}
\hbox{CALT-68-2107}
}}
{\vbox{
\centerline{Heterotic Matrix String Theory}}}

\centerline{David A. Lowe}
\medskip
\centerline{California Institute of Technology}
\centerline{Pasadena, CA  91125, USA}
\centerline{\tt lowe@theory.caltech.edu}
\bigskip
M-theory suggests the large $N$ limit of the matrix description of a
collection of $N$ Type IA D-particles should provide a nonperturbative
formulation of heterotic string theory. In this paper states
in the matrix theory corresponding to fundamental heterotic strings
are identified, and their interactions are studied. Comments are made
about analogous states in Type \IIA\ string theory, which correspond
to bound states of D-particles and D-eightbranes.

\Date{April, 1997}

\lref\sen{A. Sen, ``A Note on Marginally Stable Bound States in Type
II String Theory,'' \PR {\bf D54} (1996) 2964; hep-th/9510229.}
\lref\daniel{U. Danielsson and G. Ferretti, ``The Heterotic Life of
the D-particle,'' hep-th/9610082.}
\lref\polwit{J. Polchinski and E. Witten, ``Evidence for
Heterotic-Type I String Duality,'' \NP {\bf B460} (1996) 525,
hep-th/9510169.}
\lref\kacsil{S. Kachru and E. Silverstein, ``On Gauge Bosons in the
Matrix Model Approach to M Theory,'' hep-th/9612162.}
\lref\danieltwo{U. Danielsson, G. Ferretti and B. Sundborg,
``D-particle Dynamics and Bound States,'' \IJMP {\bf
A11} (1996) 5463, hep-th/9603081.}
\lref\bfss{T. Banks, W. Fischler, S. Shenker and L. Susskind,
``M-Theory as a Matrix Model: A Conjecture,'' hep-th/9610043.}
\lref\horava{P. Horava and E. Witten, ``Heterotic and Type I String
Dynamics from Eleven Dimensions,'' \NP {\bf B460} (1996) 506,
hep-th/9510209.}
\lref\dai{P. Horava, ``Strings on Worldsheet Orbifolds,'' \NP {\bf B327}
(1989) 461; J. Dai, R.G. Leigh and J. Polchinski, ``New Connections
Between String Theories,'' \Mod {\bf A4} (1989)
2073;  P. Horava, ``Background Duality of Open-String Models,''
\PL {\bf B231} (1989) 251.}
\lref\gresch{M. Green, J. Schwarz and E. Witten, ``Superstring
Theory,''
Cambridge University Press, 1987.}
\lref\oldzero{M. Claudson and M.B. Halpern, ``Supersymmetric Ground
State Wave Functions,'' \NP {\bf B250} (1985) 689;
M. Baake, P. Reinicke and V. Rittenberg, ``Fierz Identities for Real
Clifford Algebras and the Number of Supercharges,'' \JMP {\bf 26} (1985) 1070;
R. Flume, ``On Quantum Mechanics with Extended Supersymmetry and
Nonabelian Gauge Constraints,'' \AP {\bf 164} (1985) 189.}
\lref\doug{M.R. Douglas, D. Kabat, P. Pouliot and S.H. Shenker,
``D-branes and Short Distances in String Theory,'' \NP {\bf B485}
(1997) 85, hep-th/9608024.}
\lref\kabat{D. Kabat and P. Pouliot, ``A Comment on Zerobrane Quantum
Mechanics,'' \PRL {\bf 77} (1996) 1004,
hep-th/9603127.}
\lref\hoppe{B. de Wit, J. Hoppe, H. Nicolai, ``On the Quantum
Mechanics of Supermembranes,'' \NP {\bf B305} (1988)
545.}
\lref\lowe{D.A. Lowe, ``Bound States of Type I$'$ D-particles and
Enhanced Gauge Symmetry,'' hep-th/9702006.}
\lref\dbranes{J. Polchinski, ``TASI Lectures on D-Branes,''
hep-th/9611050; J. Polchinski, S. Chaudhuri and C. Johnson,
``Notes on D-Branes,'' hep-th/9602052.}
\lref\duality{E. Witten, ``String Theory Dynamics in Various
Dimensions,'' \NP {\bf B443} (1995) 85, hep-th/9503124;
C. Hull and P. Townsend ``Unity of Superstring Dualities,'' \NP {\bf
B438} (1995) 109, hep-th/9410167.}
\lref\mtheory{J. Schwarz, ``The Power of M-Theory,'' \PL {\bf B367}
(1996) 97, hep-th/9510086.}
\lref\green{M. Green and J. Schwarz, ``The hexagon gauge anomaly in
Type I superstring theory,'' \NP {\bf B255} (1985) 93.}
\lref\bss{T. Banks, S. Seiberg and E. Silverstein, ``Zero and
One-dimensional Probes with N=8 Supersymmetry,'' hep-th/9703052.}
\lref\moore{R. Dijkgraaf, G. Moore, E. Verlinde and H. Verlinde,
``Elliptic Genera of Symmetric Products and Second Quantized
Strings,'' hep-th/9608096.}
\lref\dvv{R. Dijkgraaf, E. Verlinde and H. Verlinde, ``Matrix String
Theory,'' hep-th/9703030.}
\lref\banksei{T. Banks and N. Seiberg, ``Strings from Matrices,''
hep-th/9702187.}
\lref\motl{L. Motl, ``Proposals on Nonperturbative Superstring
Interactions,'' hep-th/9701025.}
\lref\witbound{E. Witten, ``Bound States of Strings and p-Branes,''
\NP {\bf B460} (1996) 335,
hep-th/9510135.}
\lref\papadop{G. Papadopoulos and P.K. Townsend,  ``Kaluza-Klein on
the Brane,'' \PL {\bf B393} (1997) 59, hep-th/9609095.}
\lref\banks{T. Banks and L. Motl, ``Heterotic Strings from Matrices,''
hep-th/9703218.}
\lref\srey{N. Kim and S.J. Rey, ``M(atrix) Theory on an Orbifold and
Twisted Membrane,'' hep-th/9701139.}
\lref\dougbrane{M.R. Douglas, ``Branes within Branes,''
hep-th/9512077.}
\lref\itoyama{Y. Arakane, H. Itoyama, H. Kunitomo and A. Tokura, 
``Infinity Cancellation, Type I$'$ Compactification and 
String S-Matrix Functional,'' hep-th/9609151.}

\newsec{Introduction}

D-brane techniques \dbranes\ provide us with new ways to study the
nonperturbative dynamics
of  string theory.  One may use these methods to test the web of
strong/weak  coupling dualities
that have been proposed in recent years \duality.
Typically, these dualities have their simplest interpretation
when the theories are viewed as compactifications of a hypothetical
eleven-dimensional
theory known as M-theory \mtheory. D-brane techniques therefore yield much
information about the structure of
M-theory. In fact,
it has been conjectured the full dynamics of  M-theory with the 11th
dimension decompactified may be
described by the large $N$ matrix quantum mechanics of a system of $N$
D-particles of Type \IIA\
string theory \bfss.
One can carry the same ideas over to the compactification of M-theory
on
$S^1/\IZ_2 $ which
has been conjectured to describe the strongly-coupled dynamics of the
$E_8\times E_8$
heterotic string \horava, and attempt to describe the dynamics of
this system using
the large $N$ limit of the matrix quantum mechanics of
Type IA D-particles\foot{In this paper we will use the notation Type IB
to refer to the usual Type I string theory, and Type IA to refer to
the theory obtained by T-dualizing Type I on $S^1$ \dai, which has
previously been referred to as Type I$'$ string theory.}
\refs{\daniel \kacsil \lowe {--} \srey}.

In this paper, we will continue our study of this system \lowe,
elaborating on
the spectrum
of states and their interactions. Applying a T-duality transformation
along
the $S^1/\IZ_2$
direction, the supersymmetric quantum mechanics is recast into the form
of a gauge theory in two dimensions with $(0,8)$ supersymmetry. The coupling
constant of this gauge theory $g$ scales  with the length, thus
in the infrared this gauge theory is expected to flow to some
nontrivial
conformal
field theory, which we identify as a $S_N$ orbifold heterotic sigma-model.
The spectrum of the heterotic string is recovered in this formulation, and
the interactions of these states are studied. One may also consider the
limit in which $g$ is held fixed and one treats the gauge potential
as a slowly varying degree of freedom in a Born-Oppenheimer
approximation. The states required by equivalence with M-theory
on $S^1/\IZ_2$ duly appear in this limit.
We conclude with some comments
about analogous bound states of D-branes in Type \II\ string theory.

\newsec{Matrix Field Theory}

Our starting point will be the two-dimensional
gauge theory describing the light excitations
of $N$ coincident D-strings in Type IB. This system is T-dual to the
system of
D-particles in
Type IA studied in \refs{\daniel \kacsil {--}\lowe}.
The relation with $E_8 \times E_8$
heterotic string theory compactified on a circle (the 9 direction)
may be seen by the following chain of
dualities.
\eqn\diagra{
\matrix{ {\rm IA~ D-particle~ \#} & \maphor{{\rm T-duality}} &
{\rm IB~ D-string~ winding~ \#} \cr
\mapver{9-11~ {\rm flip}} & & \mapver{{\rm S-duality}} \cr
E_8\times E_8~ {\rm Heterotic} ~p_+ & \maphor{{\rm T-duality}} &
SO(32)~{\rm Heterotic~ winding~ \#} \cr}
}

The two-dimensional gauge theory
is obtained from the massless
sector of open strings ending on D-strings and D-ninebranes,
using conventional D-brane techniques \dougbrane. The interactions of these
fields are fixed by the $(0,8)$ supersymmetry. The answer one obtains is
the $O(N)$ gauge theory with action
\eqn\gaugeac{
\eqalign{
S &= {1\over 2\pi} \int {\rm Tr} \biggl( (D_\mu X)^2 -
i \theta_+^T D_-\theta_+ +
g_s^2 F^2 -i g_s^2 \lambda_- D_+ \lambda_-  + \cr &
2 i \theta_+ \lambda_- \gamma_i X^i-i\chi D_+ \chi +
{1\over g_s^2} [X^i, X^j]^2  \biggr)~, \cr}
}
in units where $\alpha'=1$, $g_s$ is the string coupling
for $E_8\times E_8$ heterotic strings, and the $\gamma_i$ are defined
as in \gresch.
Here the eight scalars $X^i$ and their superpartners, the
right-moving
fermionic fields $\theta_+$ lie in the symmetric rep of
$O(N)$. This theory has a $Spin(8)$ R-symmetry group corresponding
to the group of rotations in spacetime. The $\theta_+$ fields
transform
as the ${\bf 8_c}$ while the $X^i$ transform as the ${\bf 8_v}$.
The gauge multiplet lies in the adjoint of $O(N)$ and
consists of a gauge boson $A_\mu$ and a set of
eight left-moving fermionic fields
$\lambda_-$ which transform in the ${\bf 8_s}$. The $\chi$ fields
are purely left-moving and transform as the fundamental rep of $O(N)$
and in the fundamental of the
$SO(32)$ group of spacetime gauge symmetry. The necessity of
including
the $\chi$ fields may be seen either from the point of view of
spacetime
anomaly cancellation in Type I string theory \refs{\green,\dai,\itoyama},
where a background of
$32$ ninebranes is required, giving rise to the $\chi$ fields as 1-9
open
strings, or from the point of view of anomaly cancellation in
the two-dimensional gauge theory \bss.

It is natural to conjecture that this theory provides a
nonperturbative definition of uncompactified heterotic string theory.
However, we know from the work of Polchinski and Witten \polwit\ that
Type IB perturbation theory, on which this description is based,
will typically break down as the radius of
the
$S^1$ is varied due to dilaton and graviton tadpoles not cancelling
winding number by winding number. The only way to avoid this
breakdown of perturbation theory is to choose a Wilson line for the
$Spin(32)/\BZ_2$ degrees of freedom corresponding to
\eqn\wilsonline{
W=(({1\over 2})^8, 0^8)~,
}
in
standard notation. In Type IA language this corresponds to
putting eight D-eightbranes on each of the orientifold planes. The
$SO(32)$ gauge symmetry is broken to $SO(16)\times SO(16)$. The
Wilson line splits the $\chi$ degrees of freedom into two sets
of sixteen, which will have boundary conditions differing by a
sign as one goes around the $S^1$ direction.

We have expressed the action \gaugeac\ in terms of the string coupling
of the heterotic string. This is related to the gauge theory
coupling by $g_s = 1/g$ and $g_s$ scales inversely with the worldsheet
length
scale. The perturbative heterotic string theory should therefore
be recovered in the infrared limit. In this limit the
gauge theory should flow to some nontrivial superconformal fixed
point. We conjecture this superconformal field theory is the
heterotic sigma model with $(0,8)$ supersymmetry based on the orbifold
theory
\eqn\targsp{
 (\BR^8)^N/( S_N \sdtimes (\BZ_2)^N)~,
}
where the $S_N$ acts by permuting the $\BR^8$, their fermionic
partners and gauge fermion degrees
of freedom. The $\BZ_2$'s act by reflecting the different
components
of the $O(N)$ vectors.
In the free-string limit the space of states
will be shown to correspond to the Fock space of second-quantized
heterotic strings.

\newsec{Spectrum of States}

\subsec{Heterotic Strings}

The techniques of \refs{\moore \dvv \banksei {--} \motl} carry over
straightforwardly
to the
present case. The Hilbert space of the $S_N$ orbifold is decomposed
into
twisted sectors for each of the conjugacy classes $[g]$ of $S_N$
\eqn\hilbert{
\CH = \oplus_{[g]} \CH_{[g]}~.
}
These
conjugacy classes may be represented as cycle decompositions and then the
Hilbert spaces $\CH_{[g]}$ may be further decomposed as symmetric
products
of Hilbert spaces $\CH_{(n)}$ where $\CH_{(n)}$ is the $\BZ_n$ invariant
subspace of a single heterotic string on $\BR^8 \times S^1$ with
winding
number $n$ and Wilson line $W$. Modding out by the additional
$\IZ_2$'s is achieved by extra GSO projections acting on the
gauge fermions.

This space of states is represented by $n$
copies of the fields $x^i_I(\sigma)$ and $\chi^r_I(\sigma)$ with
$I=1,\cdots,n$ and
$\sigma \in [0, 2\pi]$, with the cyclic boundary conditions
\eqn\cycbound{
\eqalign{
x^i_I(\sigma+2\pi) &= x^i_{I+1}(\sigma) \cr
\chi^r_I(\sigma+2\pi) &= \epsilon_I \CO_W \chi^r (\sigma) ~,\cr}
}
where $\epsilon_I=\pm 1$ corresponds to reflection elements of $O(N)$
and $\CO_W$ is the action
of the Wilson line $W$ on the fermions $\chi$ which is simply
multiplication
by $-1$ for $r=1\cdots 16$ and $+1$ for $r=17,\cdots, 32$. 
These
fields may be combined into a single set of fields $x^i(\sigma)$ and
$\chi^r(\sigma)$ living on
$\sigma \in [0, 2\pi n]$ where
\eqn\combou{
\eqalign{
x(\sigma+2\pi n) &= x(\sigma) \cr
\chi^r(\sigma+2\pi n) &= \epsilon (\CO_W)^n \chi^r(\sigma)~,\cr}
}
and $\epsilon= \pm 1$. 
We label the sectors with different boundary
conditions on the $\chi$'s by $P$ for periodic and $A$ for
antiperiodic. Since the first group of 16 $\chi$'s can have different
boundary conditions to the second group, we need one label for
each group.  For $n$ even, we have the
$PP$ and $AA$ sectors of the usual heterotic string, while
$n$ odd gives the $AP$ and $PA$ sectors. The vacuum energy vanishes in the
$AP$ and $PA$ twisted sectors, while in the twisted sector with $PP$
boundary conditions the left-movers make a nonzero contribution equal
to $1/n$, and in the $AA$ sector $-1/n$. When we rescale 
$L_0 +\bar L_0$ by a factor of $n$ so it is canonically
normalized with respect to $x^i(\sigma)$ the vacuum energy is $1$ in
the $PP$ sector and $-1$ in the $AA$ sector as
expected for a single heterotic string.
The usual
$E_8\times E_8$
heterotic string
has two GSO projections consisting of keeping states invariant
under $(-1)^{F_1}$ and $(-1)^{F_2}$, where $(-1)^{F_1}$ anticommutes
with the first group of $\chi^r$ and commutes with the second,
and vice-versa for $(-1)^{F_2}$. The $Spin(32)/\IZ_2$ heterotic theory
on the other hand has just a single GSO projection consisting of
$(-1)^{F_1+F_2}$. This projection is to be identified with
a $\BZ_2$ element of $O(N)$ which acts as $-1$
on vectors $\chi^r_I$.

Fundamental heterotic strings are obtained in a large $N$ limit,
by considering the twisted sector for some cycle of length $n$
with $n/N$ finite. This corresponds to considering a string carrying
a finite longitudinal momentum $p_+ = n/N$.
Invariance under $\BZ_n$ implies that $L_0 - \bar L_0$ is a multiple
of $n$. The mass of such a state diverges as $N\to \infty$ unless
$L_0-\bar L_0$ acting on
the state vanishes. In this way, the usual level-matching condition of
the heterotic string is recovered.

What happened to the second GSO projection of the $E_8 \times E_8$
heterotic string? The theory we have constructed describes
$E_8\times E_8$ heterotic strings
compactified on some large $S^1$ with $n$ quanta of Kaluza-Klein
momentum present.
This is equivalent via T-duality
to an $Spin(32)/\IZ_2$ heterotic string with winding number $n$
compactified on a small $S^1$ with
the Wilson line $W$ present. The worldsheet fields that
emerge from the sigma model are naturally identified with these
fields, hence only one GSO projection appears. To take the large
$N$
limit it is convenient to $T$-dualize the sigma model fields, which
introduces the second GSO projection, as in the usual $E_8 \times E_8$
theory.
We see therefore that the Hilbert space $\CH$
corresponds to the second-quantized Fock space of free
$E_8\times E_8$ heterotic strings.

\subsec{Type IA BPS Bound States}

When the right-movers of the heterotic strings are in their ground
states, the states found above will be BPS
saturated and using the 9-11 flip duality \diagra\ may
be reinterpreted as BPS bound states of Type IA D-particles
(and tensor products of such states). The states we will be interested
in here break half of the spacetime supersymmetry and
correspond to bound states at
threshold of Type IA D-particles.
The infrared limit of the gauge theory amounts to
considering the Born-Oppenheimer approximation in the supersymmetric
quantum mechanics of D-particles on the branch of the
moduli space when they are stuck to one of the orientifold planes.
The above results for the BPS heterotic string states found above
carry over immediately to this case. The BPS bound
states of $N$ D-particles are to be identified with the massless
states in the $\IZ_N$ twisted sector. For $N$ even one finds
the adjoint of $SO(16)\times SO(16)$ in the ${\bf 8_v +
8_s}$,
and a gauge singlet in the ${\bf 8_v\times (8_v + 8_s)}$ coming from
the $AA$ sector, while
for $N$ odd the ${\bf (1,128)+(128,1)}$ in the ${\bf 8_v +8_s}$,
is found in the $AP$ and $PA$ sectors.

One may also consider the branch when the D-particles move off of the
orientifold plane. On this branch of the moduli space the scalar $A$
coming
from the gauge field up in two dimensions is treated as a slow
variable
in the Born-Oppenheimer approximation. For the moment, let us focus on
the case of two D-particles considered in \lowe. There are two
distinct limits to be considered. When $l_{11} \ll A \ll l_s$,
where $l_s$ is the string length, the relevant quantum mechanics is
obtained by dimensionally reducing the $N=2$ system in two dimensions.
Different
boundary conditions for the $\chi$ fields in two
dimensions give rise to different
sectors of the SUSY quantum mechanics with different Hamiltonia.
In the sector where all the $\chi$ fields satisfy
antiperiodic boundary conditions the states found \lowe\  included
the adjoint of $SO(16)\times SO(16)$ in the
\eqn\vecst{
{\bf 8_v +8_s}}
and a gauge singlet in the
\eqn\gravst{
{\bf 8_v\times (8_v + 8_s)}~.
}
The quantum
numbers of these states
match the massless states coming from the $AA$ sector of the CFT discussed
above.
This is a necessary condition
for the exact wavefunctions to smoothly match as one moves from
one branch of the moduli space to the other. The one-loop
corrections to the effective kinetic term for $A$ \daniel\ are of the
form $g_{IA} {\dot A}^2/A^3$, indicating that corrections to this
approximation
set in at the eleven-dimensional Planck scale $l_{11}$.
For $l_{11} \ll A \ll l_s$ the wavefunction of these states should
vary as $c_1 A + c_2$ where $c_1$ and $c_2$ are constants, which
can in principle be determined by matching onto the behavior as $A\to
l_{11}$ and as $A\to l_s$. We know these states should spread onto the
branch of moduli space at $A=0$, so we expect $c_2$ to be nonzero.
In the limit $A \gg l_s$, the dominant interaction is closed string
exchange, and we expect the Hamiltonian to flow to that describing
a Type IIA D-particle, interacting with its mirror image \daniel.
The $\chi$ modes will become massive, and completely decouple from
this Hamiltonian. The zero energy states will correspond to a
singlet of the gauge group in the ${\bf (8_v+8_s)\times (8_v+8_c)}$ of
$Spin(8)$ which are
the quantum numbers of the ground state
of a single Type \IIA\ D-particle. This state does not match onto
the charged states found for $A \ll l_s$, so the wavefunction of \vecst\
should vanish when $A \gg l_s$.
In this way, we see the states \vecst\ are localized
near the orientifold plane.

In the sector when all the $\chi$ fields satisfy periodic boundary
conditions one finds all states become massive, and this sector
decouples from the low-energy description of the D-particles.
When half of the $\chi$'s satisfy periodic boundary conditions,
and half are antiperiodic, a similar calculation to the one in \lowe\
shows the normal-ordering terms
appearing in the term in the Hamiltonian linear in $A$ vanish. In
particular,
the $\chi$ fields become massive and may be integrated out leaving
states
that are singlets under $SO(16)\times SO(16)$.
The states found in the Born-Oppenheimer approximation arise from
quantizing the fermion zero modes arising from the trace part of
$\theta_+$
and the $\lambda_-$, the left-moving
superpartners of $A$. This gives rise to the
${\bf (8_v+8_s)\times (8_v+8_c)}$ of $Spin(8)$ which are
the quantum numbers of the ground state
of a single Type \IIA\ D-particle - as expected from duality with
M-theory. Note in this case the Hamiltonian describing these degrees
of freedom is the same for all $A \gg l_{11}$. These states are not
localized near the orientifold planes.

Let us now generalize these results to arbitrary numbers of
D-particles.
For $N$ even, we can choose $A$ to break
$O(N) \to U(1)\times SU(N/2)$.
The $U(1)$ component of $A$ may be treated in a similar way to the
$N=2$ case already discussed. In the sector where the
$\chi$'s satisfy $AP$ or $PA$ boundary conditions,
the other light degrees of freedom
correspond to the $SU(N/2)$ quantum mechanics of Type \IIA\
D-particles with fixed center of mass.
The Hamiltonian for these slow modes computed in the Born-Oppenheimer
approximation matches that of $N/2$ Type \IIA\ D-particles. This
is to be expected from the M-theory point of view -- far from the
ends of interval, the states should match that of
uncompactified M-theory. In section 5 we will
comment on finding BPS bound states for this system.
In the sector where the $\chi$'s obey $AA$ boundary conditions the
$U(1)$ component of $A$ behaves as for the $N=2$ case, and again the
other degrees of freedom again take the form of the $SU(N/2)$ quantum
mechanics
of $N/2$ Type \IIA\ D-particles. Assuming this $SU(N/2)$ quantum
mechanics yields one $L^2$ normalizable bound state, one finds BPS
bound states which match onto the charged
states appearing on the other branch of the
moduli space (i.e. the adjoint and gauge singlet states)
and will be peaked near the orientifold plane by the argument given above.
Finally when $\chi$'s obey $PP$ boundary conditions all states become
massive and hence non-BPS.

For $N=1$ the D-particle is always stuck to the
orientifold, and the BPS states are just the massless states of
the CFT found in the previous section.
Only $AP$ and $PA$ boundary conditions are possible for this case and
as mentioned before,
one finds the ${\bf (1,128)+(128,1)}$ of
$SO(16)\times SO(16)$. For general $N>1$ odd, $A$
can be chosen to break $O(N) \to \IZ_2 \times U(1)\times
SU((N-1)/2)$.
This branch of the moduli space describes one D-particle
stuck on one of the orientifold planes, and $(N-1)/2$ off of the orientifold.
The slow modes consist of the $\chi$ modes arising from open
strings running between D-eightbranes and the D-particle on the
orientifold
plane,  and modes identical to $(N-1)/2$ Type \IIA\ D-particles. The
$\chi$ zero modes imply such states will always carry the
spinor charges of $SO(16)\times SO(16)$. The quantum numbers of these
match the $N$ odd states on the other branch discussed above.
The $U(1)$ component of the gauge multiplet will behave in the
same way as the $N=2$ case above, and the
wavefunctions of these states will be
localized near the orientifold planes.

\newsec{Heterotic String Interactions}

The two-dimensional gauge theory \gaugeac\ in the infrared limit gives
rise to the Fock space of free heterotic strings. Interaction terms
should correspond to irrelevant perturbations of the orbifold sigma
model. Dijkgraaf, Verlinde, Verlinde \dvv\ considered such terms
in the Type \II\ case, and much of their argumentation applies here.
The idea is that the interactions correspond to transpositions
of the $x_I$ eigenvalues when they coincide. In the heterotic case
one must also match the gauge fermion degrees of freedom, and
transpose these at the same time.
This will give rise to splitting and joining
interactions of the long strings
discussed above.

For the right-moving sector, the construction of the interaction
vertex
precisely mimics that of \dvv. We consider two eigenvalues $x_1$ and
$x_2$ which are interchanged under a $\IZ_2$ twist. In terms
of the linear combinations $x_\pm = x_1 \pm x_2$, the $\IZ_2$ flips
the sign of $x_-$ (and likewise for the right-moving
fermionic modes $\theta$).
This $\IZ_2$ orbifold is well-known, and the twist operators are
defined
by the operator product relations
\eqn\opes{
\eqalign{
\partial x_-^i(z) \cdot \sigma(0) &\sim z^{-\half} \tau^i(0) \cr
\theta_-^a(z) \cdot \Sigma^i(0) &\sim z^{-\half} \gamma_{a \dot a}^i
\Sigma^{\dot a} \cr
\theta_-^a(z) \cdot \Sigma^{\dot a}(0) &\sim z^{-\half} \gamma_{a \dot a}^i
\Sigma^{i} ~.\cr}
}
One can then show \dvv\ that the operator
\eqn\righttw{
V_R=\tau^i \Sigma^i~,
}
 is the
unique, least irrelevant perturbation that preserves $Spin(8)$ spacetime
rotations and spacetime supersymmetry. This operator has conformal
weight
$(0, {3\over 2})$.

Now consider the left-moving sector, and bosonize the gauge fermion
modes $\chi$, to yield sixteen bosonic coordinates $x^M$, with
$M=1,\cdots,16$
living on the Cartan torus of the gauge group. We define the $\IZ_2$
eigenvectors
$x^M_\pm = x^M_1\pm x^M_2$. The twist operator
for the left-moving spacetime coordinates $x_-(\bar z)$
is constructed as above. One can also define the twist operators for
the $x^M_-$ in a similar way, but now one must remember the $x^M$ are
compact bosonic coordinates. The action $x^M_- \to - x^M_-$ has two
fixed
points, so we have two twist operators $\bar\sigma_\pm$ for each $x^M$.
We now wish to construct the least irrelevant perturbation to the CFT
preserving $Spin(8)$ rotations, spacetime supersymmetry, and gauge
invariance. We also demand that the perturbation be invariant under
$z \to e^{i \theta} z$, which means the conformal weight of the
left-moving piece should be $({3\over 2},0)$. The combination
of left-moving twist operators which satisfies these conditions is
\eqn\lefttw{
V_L=\sum \bar \sigma(0) \cdot \prod_{M=1}^{16} \bar\sigma_{s_M}^M~,
}
where $s_M=\pm$ and the sum is over all permutations of the $s_M$.

The final interaction term that appears in the sigma model is then a
sum
over pairs of eigenvalues that may be interchanged
\eqn\vint{
S_{int} = \lambda \sum_{I<J} \int d^2z (V_L \otimes V_R)_{IJ}~,
}
where $I,J = 1,\cdots,N$. The coupling $\lambda$ will scale inversely
with the worldsheet length scale, therefore $\lambda$ will scale
linearly with $g_s$. The interaction corrections contained in the
supersymmetric gauge theory \gaugeac\ therefore reproduce heterotic
string interactions to first order in $g_s$.

\newsec{Comments on Type \II\ Bound States}

\subsec{Bound states of $N$ D-particles}

We saw above that the quantum mechanics describing Type IA
D-particles far from the orientifold plane reduced to
that of a collection of Type \IIA\ D-particles. The quantum mechanics
of Type \IIA\ D-particles has previously been studied in
\refs{\danieltwo \kabat \doug \oldzero  {--} \hoppe}.
For M(atrix) theory to be correct
for the case of eleven uncompactified dimensions, these Type \IIA\
D-particles must form bound states at threshold. The prediction
is that one normalizable
bound state at threshold of $N$ D-particles appears
for every $N$, and this bound state lies in an ultra-short multiplet
of the ten-dimensional  Type \IIA\ supergravity.

If such a normalizable bound state appears in uncompactified Type \IIA,
then it should show up as a normalizable bound state when
we further compactify on a torus. We can then recast the problem
into a statement about the ground states of $d=2$ $SU(N)$ Yang-Mills
theory with $(8,8)$ supersymmetry (here we have factored out the
$U(1)$ describing the center of mass degrees of freedom).
That is, there should be one $L^2$ normalizable zero-energy
ground state which appears regardless of the boundary conditions \sen.

We can make a heuristic argument for the existence of such a state as
follows. Since the state is normalizable (in fact it should fall off
as $1/r^7$ up in ten dimensions), it should remain a bound state if
we add a perturbation that only effects the large distance behavior
of the fields. In fact, if we choose the perturbation judiciously,
the bound state at threshold should become a bound state with mass
gap.

The two-dimensional super Yang-Mills theory can
be thought of as the dimensional
reduction of a $d=4$ $\CN=4$ theory. The vector multiplet in four dimensions
can be decomposed into a single
$\CN=1$ vector multiplet and three $\CN=1$ chiral multiplets which we
denote by $X$, $Y$ and $Z$, all in the adjoint rep of the gauge group.
Let us assume the perturbation of the superpotential parametrized by
$m$
\eqn\superpt{
W= {\Tr }\biggl( X [Y,Z] - \half m (X^2+Y^2+Z^2) \biggr)~,
}
behaves in the way described. We may now use a variant of the
arguments of \witbound. Since the ground state we are
interested in now has a mass gap, we can actually treat $m$ as being
large,
and use semiclassical methods.

The semiclassical vacua correspond to critical points of the
superpotential
\superpt\ which are solutions of
\eqn\critpts{
\eqalign{
[X,Y] &= m Z \cr
[Y,Z] &= m X \cr
[Z,X] &= m Y ~.\cr}
}
These equations are just the commutation relations of $SU(2)$. There three
are classes of solutions to these equations: (i) trivial solution
for which $X=Y=Z=0$, (ii) non-trivial reducible embedding of $SU(2)$
into
$SU(N)$, (iii) irreducible embedding of $SU(2)$ into $SU(N)$.
For cases (i) and (ii) part of the gauge symmetry is unbroken and one
is left with a two-dimensional gauge theory with $(2,2)$
supersymmetry.
These
vacua will not have mass gaps due to massless fermions.
Case (iii) however leads to a solution
which completely breaks the gauge symmetry. In this case all fields
become
massive and one is left with a single bosonic ground state. Including
the
fermion zero modes coming from the center of mass degrees of freedom,
one obtains a single bound state of $N$ D-particles with the
expected spacetime quantum numbers.

\subsec{Bound states of $N$ D-particles and $M$ D-eightbranes}

Studies of solitonic solutions of Type \IIA\ supergravity do not
show any sign of supersymmetric bound states of D-particles and
D-eightbranes \papadop. We can now understand this from the point of view of
the two-dimensional supersymmetric gauge theory describing this
system. As in the Type IA case, we have a theory with $(0,8)$
supersymmetry,
this time with $U(N)$ gauge multiplet $(A, \lambda_-)$.
The matter content consists of
multiplets $(X^i, \theta_+)$ in the adjoint of $U(N)$, and
left-moving
fermions in the $(N,\bar M)+ (\bar N, M)$.
Consider the conformal field theory this will flow to in the infrared.
One will be left with the fields living on the Cartan torus of $U(N)$
corresponding to the diagonal components of $X^i$, $\theta_+$ and
$\lambda_-$ together with the other left-moving fermions. For all $N$ and $M$
this conformal field theory will not have a massless ground
state, hence no BPS bound state is possible.
Massless ground states are possible if one considers for example two
sets of separated
D-eightbranes. The condition then is that there is a choice of
boundary conditions on
the left-moving fundamental fermions which gives zero contribution to the
vacuum energy.

\bigskip

\centerline{\bf Acknowledgments}

I wish to thank M. Douglas, G. Papadopoulos, J. Schwarz, A. Sen and C. Vafa
for helpful discussions.
This work was supported in part by DOE grant DE-FG03-92-ER40701.

\bigskip
\centerline{\bf Note added}

While this manuscript was being completed the preprint \banks\
appeared in which closely related results are reported. I understand
related ideas have also been considered by R. Dijkgraaf and S.J. Rey
and by P. Horava.

\listrefs
\end